# Double Resonant Vertically Accessed Optical Waveguide Sensor

**Anat Demeter-Finzi and Shlomo Ruschin**[*]

*Department of Physical Electronics, School of Electrical Engineering Faculty of Engineering, Tel-Aviv University, Tel-Aviv 69978 Israel*

*\*Corresponding author:* ruschin@tauex.tau.ac.il

*Abstract: We present a new waveguided-sensing configuration combining the advantages of two effects, namely, two-wave interference and multiple-wave interference in a single sensing device. The sensor is accessed vertically from above by means of a three-port grating, and its readout can be accomplished both in an up-reflective or waveguided mode. The superposition of the two outcoming signals results in a variety of transmission characteristic curves depending on fabrication parameters, which are optimized in order to attain enhanced sensitivity. The sensor is analyzed by means of a simple scattering matrix model and confirmed by full FDTD simulations.*

## 1. Introduction

Optical sensors based on grating couplers and waveguides attracted considerable attention for many years owing to their easy readout, compact design and ability to detected small concentration of the sensed material [1–3]. When also interference effects are involved in the detection, enhanced sensitivity is attained. Two of the most common and highly sensitive configurations include resonators [4,5] and two-wave interferometers [6,7]. They have the ability to detect refractive index changes in the range of $10^{-6}$ RIU, and even less when dispersion effects are suitably harnessed [8]. The out-of-plane optical access option for coupling light into waveguides is particularly advantageous in biosensing, as manifested in widely deployed SPR configurations [9]. A perfect vertical coupling configuration [10,11] has further technical advantages since it reduces system complexity alignment and tuning. For that purpose, distinct optimization approaches have been recently published with specialized grating designs aimed to couple light into a single direction [12–14]. Distinct from previous works, here we suggest a new sensing configuration combining the advantages of two effects, namely, two-wave interference and multiple-wave interference in a single sensing device. The suggested sensor has small dimensions, is feasible to fabricate by standard microlithography, and simple to tune and couple light into. We analyze and optimize the proposed device using a simple analytical model [15,16] which gives the user the ability to design and achieve optimal sensitivity results straightforwardly.

The proposed sensor is firstly composed of a 3-port input grating for coupling the light vertically into an optical waveguide. The grating couples light symmetrically to the left and the right directions, and a suitable grating back-reflector is placed on one of the paths, in order to direct the light back towards the input grating, creating a Fabry-Perot (FP) cavity. The sensor works on an interference effect between two light beams: one that is coupled out of the FP resonator formed between the two gratings and the other one that is coupled directly by the input vertical grating to the waveguide into the right direction. By measuring the interference between these two unbalanced arms, we get an output signal that changes in an asymmetrical and periodical way during the sensing process. The sensor here is fully analyzed by a scattering matrix (S-matrix) formalism previously developed by us in order to optimize a waveguide grating coupler [16]. Using that model, the absolute sensitivity value of the sensor is extracted, and an optimized device is designed. The model also allows us to investigate the output intensity pattern changes for different reflection and transmission coefficients of the input vertical grating and for different values of the reflection from the Bragg back-reflector. Additionally, we simulate the sensor numerically by FDTD and compare the results with our simple analytical model, confirming the ability of that model to predict satisfactorily the intensity pattern change of the sensor during the sensing process. Finally, we compare our sensor to a standard FP resonator sensor with the same quality factor (Q-factor), showing a significant advantage in the limit of detection (*LOD*) of our sensing device.

## 2. Sensor Design and Optimization

### 2.1 Sensor design

The suggested double resonant vertically accessed (DRVA) sensor is depicted in Fig.1. It is composed of a planar optical waveguide and two gratings: One grating for coupling the light vertically into the waveguide in both directions symmetrically (right grating in Fig.1) and the other, acting as a Bragg back-reflector, (left grating in Fig.1) for guiding the light to the right direction. Between both gratings, an FP cavity is formed and at the measured output port (number 3), interference takes place between two signals: one that is directly coupled by the input grating and another that is guided into the waveguide and back-reflected at the left edge of the waveguide by Bragg reflector, after undergoing multiple reflections inside the resonator. A similar interference effect takes place for the up-reflected beam.

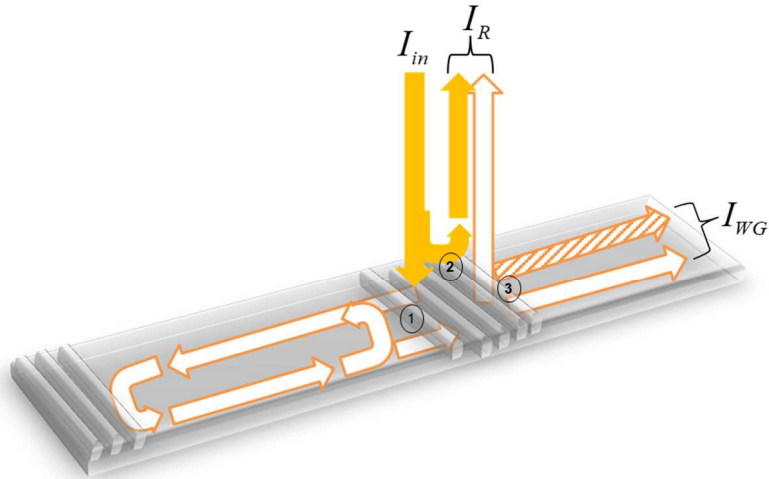

Fig.1. Schematic drawing of the suggested sensing device. The sensor is composed of two gratings: one for coupling the light vertically into the waveguide (right) and the other (left) for reflecting the light back to the right direction. The numbers represent the different ports of the input grating. $I_{in}$, $I_R$ and $I_{WG}$ are the input, up-reflected and waveguide out-coupled intensities respectively.

During the sensing process, the waveguide's effective refractive index is perturbed in the region between the gratings and as a result, the accumulated phase inside the resonator changes and the measured output intensity, $I_{WG}$, changes as well.

This design combines therefore two well-known sensitive configurations: the FP resonator and two-wave interferometer. The resonator output signal is interfering with the directly in-coupled light at the waveguide port. The combination of these two signals enhances the sensitivity furnishing an asymmetric (Fano-like) output intensity function shape.

Another possible output option of this sensor is to choose the up-reflection port, number 2 in Fig.1, as the reading port. Here interference takes place between signal that is directly up-reflected by the input grating and the resonator's output signal. This output signal has also an asymmetric behavior and at some working regimes shows even higher sensitivity. The choice about which output port to use depends on the user's system limitations and convenience. Most of the examples and formulas in the following will be given for the waveguide port (number 3), but by simple adjustments, they can be derived to analyze the up-reflected port as well.

**2.2 Sensor analysis and optimization**

We analyze the sensor's electrical fields and the measured output intensity by means of simple analytical model based on S-matrix formalism that was recently applied by us to optimize an highly efficient coupler [16] in a similar configuration.

The S-matrix model is a general model that can be applied in order to analyze any 3-port coupler-based device. Each element in the matrix represents a different reflection or transmission coefficient of the electrical field at one of the three ports of the grating:

$$S = \begin{pmatrix} \xi \exp(i\phi_\xi) & \sigma \exp(i\phi_\sigma) & \eta \exp(i\phi_\eta) \\ \sigma \exp(i\phi_\sigma) & \rho \exp(i\phi_\rho) & \sigma \exp(i\phi_\sigma) \\ \eta \exp(i\phi_\eta) & \sigma \exp(i\phi_\sigma) & \xi \exp(i\phi_\xi) \end{pmatrix} \quad (1)$$

Where $\rho, \eta, \sigma, \xi$ are the reflection and transmission magnitudes coefficients and $\phi_\rho, \phi_\eta, \phi_\sigma, \phi_\xi$ are the corresponding phases as defined in [15].

As explained in [15,16], reciprocity relations and power conservation of the system allow all the nine components of the S-matrix to be reduced to two independent variables chosen as ρ and η representing the up-reflection at port 2 and the transmission between port 1 and 3 respectively. By further assuming a steady state regime, the up-reflected field into the optical source and the waveguide right propagating intensities are calculated:

$$\begin{aligned} I_R &= |a_1 \sigma \exp(i\phi_\sigma) + a_2 \rho \exp(i\phi_\rho)|^2, \\ I_{WG} &= |a_1 \eta \exp(i\phi_\eta) + a_2 \sigma \exp(i\phi_\sigma)|^2. \end{aligned} \quad (2)$$

Where $a_1$ is the complex electrical field coefficient of the right propagating field from the back-reflector to the input vertical grating direction, and $a_2$ is the input complex electrical field amplitude ($I_{in} = |a_2|^2$).

The relation between the complex field amplitudes, $a_1$ and $a_2$, as derived in [16] furnishes the following relation:

$$a_1 = \frac{\sigma \exp(i\phi_\sigma) r_G \exp(-2ikL)}{1 - r_G(\xi \exp(i\phi_\xi)) \exp(-2ikL)} a_2, \qquad (3)$$

where $L$ is the waveguide length, $k$ is the light propagation constant inside the waveguide, and $r_G$ is the waveguide left reflector reflectivity.

After calculating the output intensities, the sensitivity and $LOD$ of the sensor are furnished:

$$Sensitivity = \frac{\partial(I_{out}/I_{in})}{\partial n_{eff}}; \quad LOD = 0.2\% \cdot \left(\frac{I_{out}/I_{in}}{Sensitivity}\right) \qquad (4)$$

Where $I_{out}$ is the measured intensity at the output port ($I_{WG}$ or $I_R$), $I_{in}$ is the input intensity and $n_{eff}$ is the waveguide effective refractive index inside the resonator. The $LOD$ criterion of resolution of 0.2% in power within a single oscillation for optical sensors was suggested in [17]. The figures of merit (FOM) in eq. (4) were defined with respect of $n_{eff}$ for standardization, and further on we provide additional calculations for changes in the cover medium.

The S-matrix model predicts the intensity variations caused by the sensing process and gives the user a practical analytical tool to optimize its sensor design. This concept together with practical simulations examples are demonstrated in the next section.

## 3. Results and Discussion

In the modeling, a source wavelength of $\lambda=1.55\mu m$ and a silicon-on-isolator waveguide of 100nm thickness and $L=100\mu m$ length were assumed. The waveguide effective refractive index was changed in the range of $2.12<n_{eff}<2.13$, where only the fundamental TE mode is excited. At the first stage, an ideal back-reflector with reflectivity close to one was assumed ($r_G=0.999$). Solving for the entire resonator and using the S-matrix model, we plotted a contour map of the sensitivity at the waveguide port as a function of $\rho$ and $\eta$ (the two independent reflection and transmission coefficients of the input vertical grating). Detailed explanation and guidelines for plotting such a map were described in [16]. In Fig.2(a) the contour map is depicted. Each point in this map represents a different input grating design, meaning a grating with different ρ and η values. Then, for each point in the contour map the waveguide effective refractive index is changed as a result of the sensing process, the waveguide output intensity is deduced, and the sensitivity is calculated. For ρ=0.99 and η=0.06 the absolute sensitivity value is 55,473 $[RIU^{-1}]$ corresponding to $LOD$ of $3.6 \cdot 10^{-8}$ $[RIU]$. As seen by this result, the theoretical sensitivity limit is achieved in the boundary region were ρ–>1 and η–>0. This point is impractical for fabrication, but its vicinity provides guideline directions to improve the sensor performance. In Fig. 2(b) the output intensity as a function of the waveguide effective refractive index is plotted at the maximal sensitivity point. The slope of the graph is very steep, reflecting the high sensitivity of that configuration. This significant sensing performance is attained for a rather short waveguide length and a proportional performance improvement will scale up by increasing it. Compared to interferometric sensors based on grating couplers [18,19], where the output intensity is sinusoidal and in order to make it steeper one needs increase the interaction length, in our sensing configuration a cavity is additionally formed between the gratings and higher sensitivities can be achieved for a shorter interacting waveguide length. When compared to a customary FP sensor, the interference of this outcoupled signal from the cavity with the directly coupled signal increases the sensitivity further as the asymmetric output intensity function shows.

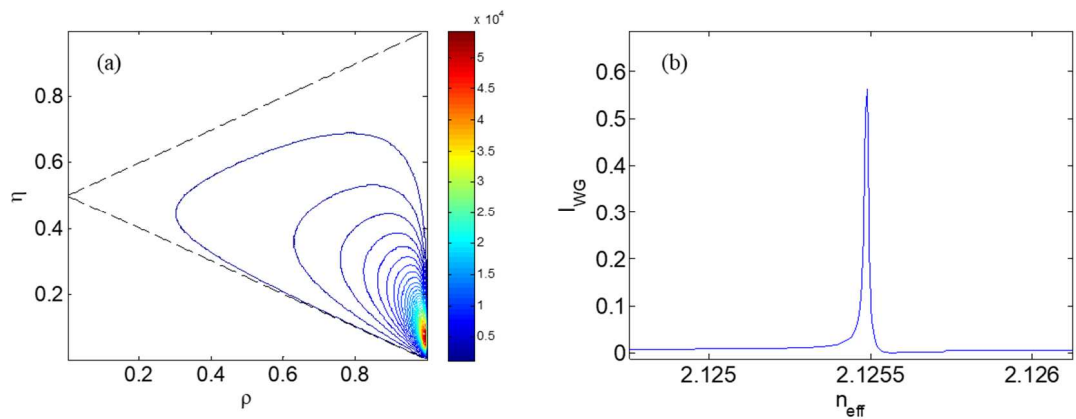

Fig.2. (a) Sensitivity contour plot of the sensor for an ideal back reflector ($r_G \to 1$). From this contour the optimized sensitivity value is extracted. (b) plot of the waveguide output intensity as a function of the waveguide effective refractive index in the most sensitive coordinate (ρ=0.99, η=0.06).

Another noteworthy property of this device is that different values of ρ and η furnish entirely different response shapes featuring improved linearity and extinction ratio, attractive for digital and analogue modulation applications [20]. In Fig.3 several plots of the intensity as function of the waveguide effective refractive index are presented. Each figure is matched to different ρ and η set of values. As seen, all graphs are periodic but have different transmission shapes. As the ρ value is increasing and η value is decreasing, the slope of the graph becomes steeper resulting in an enhanced sensitivity. The common behavior of all the graphs in Fig.3 is the asymmetric function shape which is a major advantage for sensing devices [21].

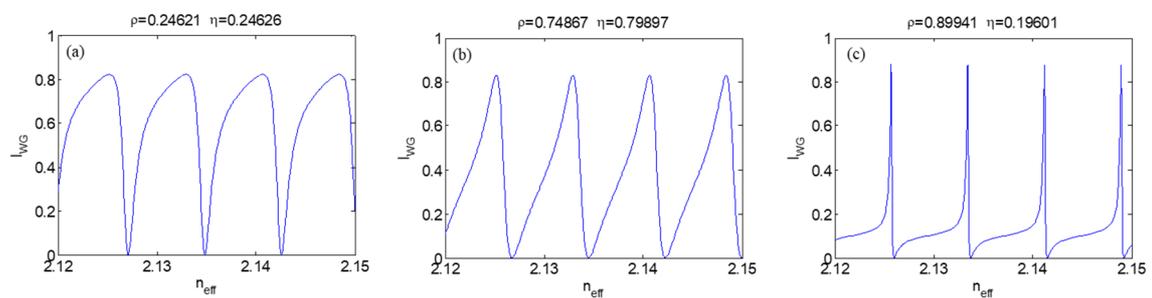

Fig.3. Sensor output intensity (port 3) as function of the waveguide effective refractive index. Each graph is matched to different ρ and η set of values: (a) ρ=0.24621 and η=0.2462, (b) ρ=0.74867 and η=0.79897 ,(c) ρ=0.89941 and η=0.19601.

Simulations insofar assumed a reflectivity close to 100% for the left-side reflector ($r_G$=0.999). In the cases of a lower value for $r_G$, the left-side reflector (Fig.1) absorbs or transmits part of the light out of the resonator and the sensitivity of the sensor decreases as seen from the graph in Fig.4(a). A worthwhile remark in that case concerns the location of the most sensitive area of the device in the contour map. As $r_G$ changes, the location of this area is displaced towards the center of the plot and the optimal sensitivity values of ρ and η change as well. In Fig.4(b) several contour maps of the sensor are presented for different $r_G$ values. From these maps we learn that in the case of $r_G$<1 the theoretical sensitivity optimum is not at a boundary region of the graph like the case in Fig.2(a), meaning the corresponding ρ and η coordinate values are also more achievable in practice.

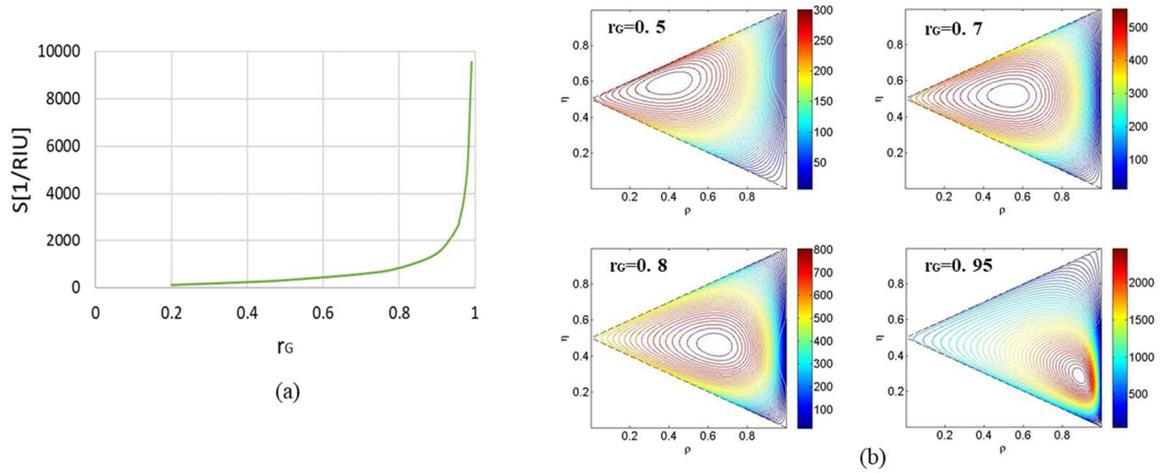

Fig.4. Sensitivity plots for different values of the Bragg reflector reflectivity: (a) plot of the sensor sensitivity as function of $r_G$, for optimal values of $\rho$ and $\eta$ for each $r_G$ value, and (b) contour plots of the device for different $r_G$ values. As $r_G$ approaches to one, the sensitivity rises and the location of the maximum sensitivity in the graph moves to the downright area of the contour map.

Until this point the results and plots given corresponded to the case of the waveguide port acting as the output port. As mentioned previously the other possible output port option, the up-reflected beam, provides similar and highly sensitive results. In Fig.5 the sensitivity of the sensor for both output ports and for different reflectivity values, $r_G$, of the Bragg back-reflector is plotted. As seen from this figure the sensitivity of the up-reflected port is higher than the sensitivity of the waveguide port in factor two or three depending on the $r_G$ value. The choice of which port to use depends on the setup limitations and user constrains as explained previously. The up-reflection port provides clearly a significant improvement in sensitivity and a simpler mechanical disposition. On the other hand, it involves back-reflection into the laser source and suitable polarization manipulation will be required in order to overcome it.

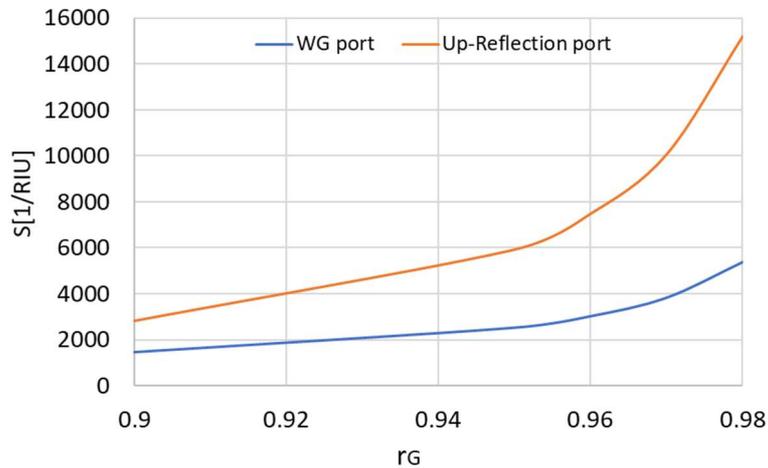

Fig.5. Graph of the sensor sensitivity for different Bragg reflector reflectivity values. The results are given for the sensor's two possible output ports: the waveguide port and the up-reflection port. As seen from the figure the up-reflection port is more sensitive by a factor of 2-3 as compared to the waveguide port.

In order to demonstrate the advantages of our sensing configuration as compared to a standard equivalent FP sensor design based on Bragg reflectors [4,5] we examined the output intensity behavior of the two devices and their *LOD* values. Both devices had the same Q-factor, output losses and resonator length. The FP sensor consists of two identical gratings in a symmetrical cavity configuration, whereas in our sensor the input grating differs in reflectivity from the back-reflector grating, i.e. the cavity is asymmetric. In Fig. 6(a) and Fig. 6(b) schematic drawings of both sensors are depicted. The performance of both sensors as a function of their common Q–factor is displayed in Fig. 6(c). Taking a specific case of a feasible reported value of $r_G$=0.95 for the back-reflecting grating [22], leads to a Q = 6399 value, translating to a *LOD* for DRVA sensor in Fig.6(a) of $7.96 \cdot 10^{-7} [RIU]$, about one third of the *LOD* of the FP sensor in Fig.6(b) ($2.46 \cdot 10^{-6} [RIU]$). These values are calculated for the generic case of changes in the effective waveguide index $n_{eff}$. For the specific case of changing the cover effective index, the *LOD* values are higher as expected $1.36 \cdot 10^{-6} [RIU]$ for the DRVA and $4.58 \cdot 10^{-6} [RIU]$ for the FP configurations respectively.

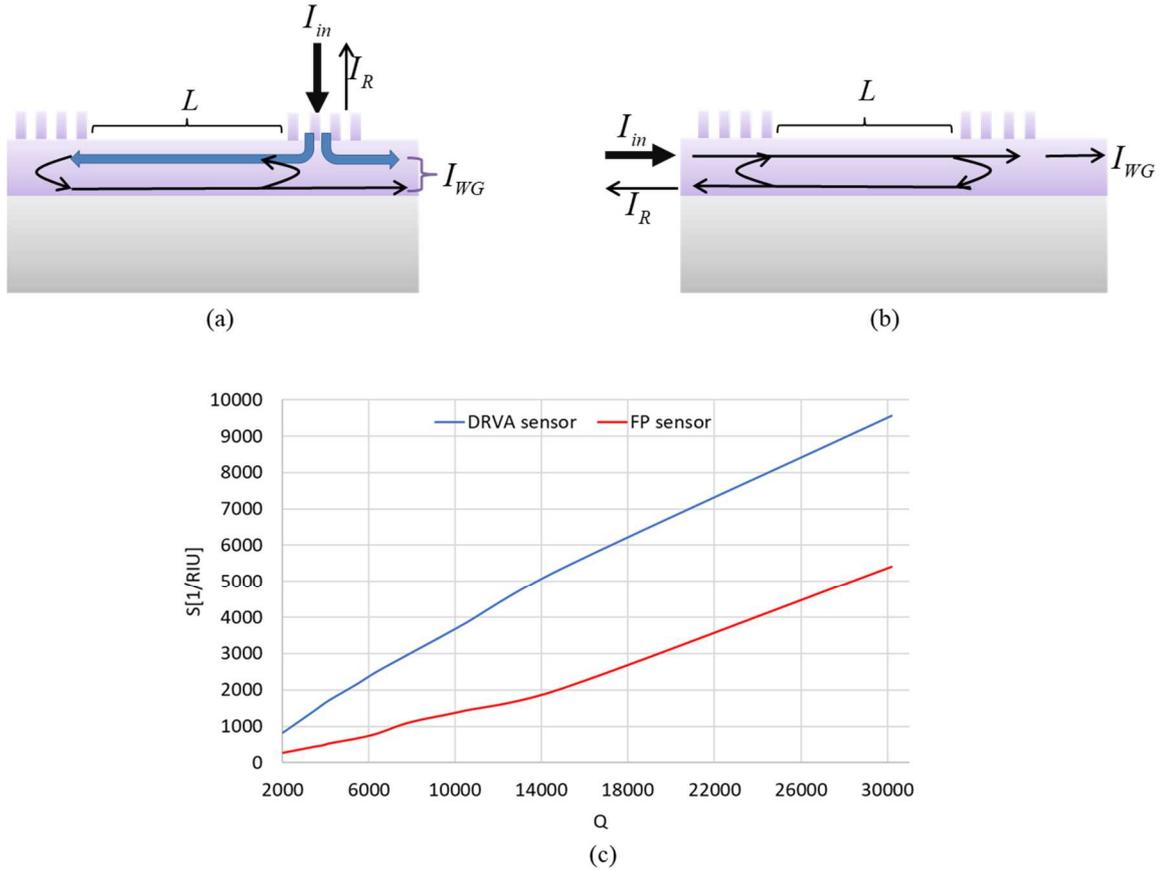

Fig.6. Illustration of the DRVA sensor design (a) and Fabry-Perot configuration (b). Both sensors have the same Q-factor and output losses. In (c) the sensitivity of both sensors is calculated as function of the sensors Q-factor. The advantage of the DRVA sensor is apparent in the whole range.

The sensitivity of the DRVA and FP as function of the sensors' Q-factor is compared in Fig.6(c). In this plot we observe a trend where our sensor sensitivity is higher than the FP sensor for the whole Q-factor range and the gap between both sensors' sensitivities is increasing at the Q-factor value is getting bigger.

As a complementary step, in order to validate our findings based insofar on a rather simple model, we performed a detailed design and FDTD numerical simulations. As shown below, we observed a high correlation between both methods. In the full numerical simulations we adopted a specific high contrast grating (HCG) scheme from the literature [23,24] with negligible light transmission to the substrate from the input grating. An ideal back-reflector was placed at the opposite end of the SOI waveguide. A much shorter waveguide length ($L$=1.5µm) was assumed in order to fit the simulator window and reduce the simulation execution time. In Fig.7(a) and Fig.7(b) the intensity as function of the waveguide effective refractive index is shown and compared with the S-matrix model predictions for two sensors designs, differing in the height of the HCG indentations (600nm and 700nm respectively), and operating at 1.55μm wavelength. The other characterization parameters of the HCG were: duty cycle=0.59 and grating period=0.733nm. In Fig.7(c) the electrical power density at the sensor region is depicted. For the comparison, we numerically evaluated ρ and η for each grating design, replaced the corresponding values of ρ and η into the S-matrix model and calculated directly the intensity as the effective refractive index changed. In the S-matrix model we assumed a reduced reflectivity of $r_G$=0.9 in order to account for losses in the input grating (radiation losses and radiation leaks to the substrate). The similarity of the waveforms attained by the two approaches is remarkable, reinforcing the reliability of the S-matrix model as a useful tool for the user to design the DRVA sensor.

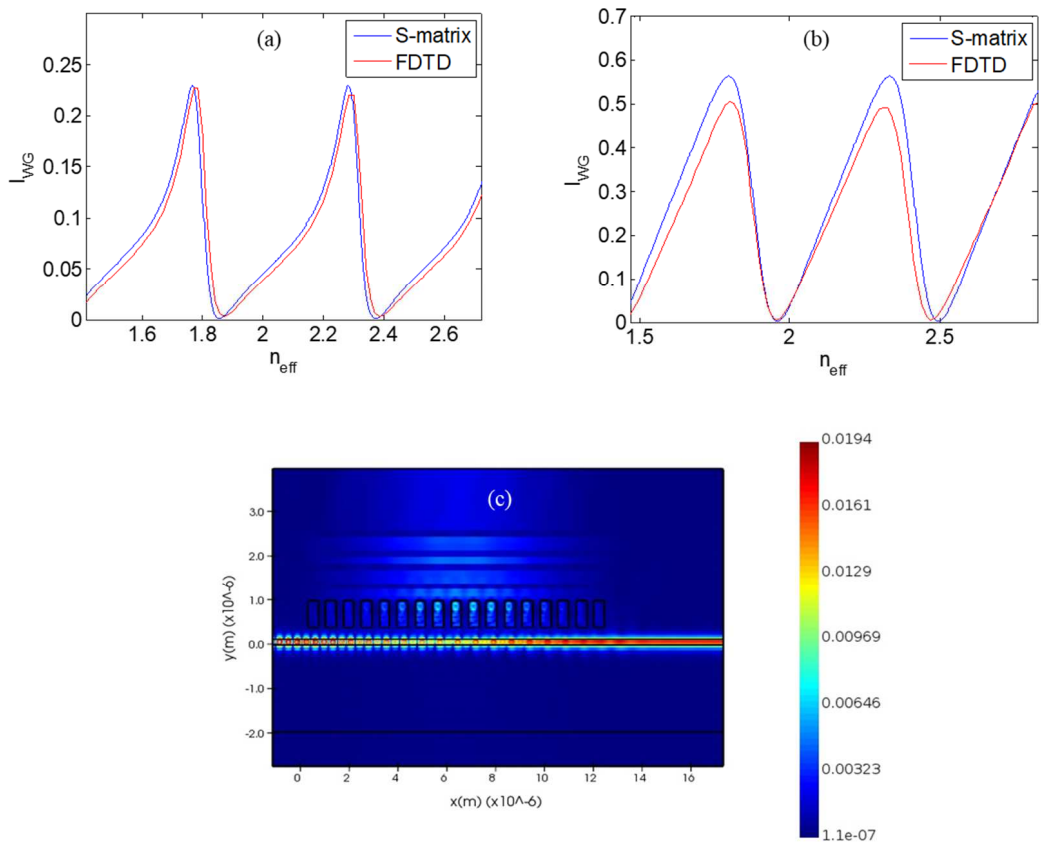

Fig.7. Comparison plots of the waveguide output intensity as function of the waveguide effective refractive index for two sensor designs with (a)600nm and (b)700nm grating indentation depths. In each of the graphs numerical FDTD and S-matrix analytical results are presented. In (c) the electrical field power density of the sensor at the input coupler vicinity section is shown.

## 4. Conclusions

We introduced a new sensing concept and device combining two resonant effects, with the additional advantage of vertical optical access. The combined resonant effects enhances the sensitivity of the device, resulting in an asymmetric characteristic curve of the output intensity as a function of the effective index of the waveguide inside the resonator. The key component of the device is a three-port diffraction grating allowing vertical optical access and the two-directional coupling into the waveguide. The modeled device comprised additionally a standard SOI waveguide and an end reflector to enclose the resonant cavity. A 100-micron sensor length was assumed, that could be scaled-up in order to enhance further the sensitivity and lower the *LOD*. Two access configurations were considered in which the input access was vertical: in the first, the output channel is within the waveguide and the second where the output channel is the upcoming counter-propagating beam relative to the incoming beam. The second option displays higher sensitivity and is simpler to interface mechanically but needs optical isolation in order to avoid damage to the laser source. We analyzed the device by means of a rather simple S-matrix model, and confirmed results by a full FTDT simulation